\documentclass[conference,final,twoside]{IEEEtran}

\usepackage{lscape}

\usepackage{xcolor}
\definecolor{codecolor}{rgb}{0.94, 1.0, 1.0}
\definecolor{commentcolor}{rgb}{0.54, 0.17, 0.89}
\definecolor{keywordcolor}{rgb}{0.28, 0.02, 0.03}
\definecolor{stringcolor}{rgb}{0.0, 0.13, 0.28}

\usepackage{listings}

\usepackage{relsize}

\newcommand{\codesize}{\smaller[1.5]}
\lstdefinestyle{Cfamily}{%
basicstyle=\ttfamily\codesize,
commentstyle=\color{commentcolor},
moredelim=[is][{\btHL[onslide=<2->{fill=red!30}]}]{/*!*/}{/*!*/},
moredelim=[is][{\btHL[onslide=<1->{fill=green!30}]}]{/*=*/}{/*=*/},
keywordstyle=\color{keywordcolor},
identifierstyle=,
stringstyle=\color{stringcolor},
showstringspaces=false
}

\lstdefinestyle{inlineCfamily}{%
basicstyle=\ttfamily,
commentstyle=\color{commentcolor},
keywordstyle=\color{keywordcolor},
identifierstyle=,
stringstyle=\color{stringcolor},
showstringspaces=false,
}

\newcommand{\code}[1]{\lstinline[language=C,style=inlineCfamily]@#1@}

\lstnewenvironment{Ccode}{\lstset{
language=C,
style=Cfamily,
numbers=right,
numbersep=5pt,
frame=single,
linewidth=8cm,
aboveskip=10pt,
belowskip=10pt
}}{}

\usepackage{xstring}

\newcommand{\MCId}[1]{%
\IfStrEqCase{#1}{%
{D1.1}{Dir~1.1}
{D2.1}{Dir~2.1}
{D3.1}{Dir~3.1}
{D4.1}{Dir~4.1}
{D4.2}{\textit{Dir~4.2}}
{D4.3}{Dir~4.3}
{D4.4}{\textit{Dir~4.4}}
{D4.5}{\textit{Dir~4.5}}
{D4.6}{\textit{Dir~4.6}}
{D4.7}{Dir~4.7}
{D4.8}{Dir~4.8}
{D4.9}{Dir~4.9}
{D4.10}{Dir~4.10}
{D4.11}{Dir~4.11}
{D4.12}{Dir~4.12}
{D4.13}{\textit{Dir~4.13}}
{D4.14}{Dir~4.14}
{R1.1}{Rule~1.1}
{R1.2}{\textit{Rule~1.2}}
{R1.3}{Rule~1.3}
{R1.4}{Rule~1.4}
{R2.1}{Rule~2.1}
{R2.2}{Rule~2.2}
{R2.3}{\textit{Rule~2.3}}
{R2.4}{\textit{Rule~2.4}}
{R2.5}{\textit{Rule~2.5}}
{R2.6}{\textit{Rule~2.6}}
{R2.7}{\textit{Rule~2.7}}
{R3.1}{Rule~3.1}
{R3.2}{Rule~3.2}
{R4.1}{Rule~4.1}
{R4.2}{\textit{Rule~4.2}}
{R5.1}{Rule~5.1}
{R5.2}{Rule~5.2}
{R5.3}{Rule~5.3}
{R5.4}{Rule~5.4}
{R5.5}{Rule~5.5}
{R5.6}{Rule~5.6}
{R5.7}{Rule~5.7}
{R5.8}{Rule~5.8}
{R5.9}{\textit{Rule~5.9}}
{R6.1}{Rule~6.1}
{R6.2}{Rule~6.2}
{R7.1}{Rule~7.1}
{R7.2}{Rule~7.2}
{R7.3}{Rule~7.3}
{R7.4}{Rule~7.4}
{R8.1}{Rule~8.1}
{R8.2}{Rule~8.2}
{R8.3}{Rule~8.3}
{R8.4}{Rule~8.4}
{R8.5}{Rule~8.5}
{R8.6}{Rule~8.6}
{R8.7}{\textit{Rule~8.7}}
{R8.8}{Rule~8.8}
{R8.9}{\textit{Rule~8.9}}
{R8.10}{Rule~8.10}
{R8.11}{\textit{Rule~8.11}}
{R8.12}{Rule~8.12}
{R8.13}{\textit{Rule~8.13}}
{R8.14}{Rule~8.14}
{R9.1}{\textbf{Rule~9.1}}
{R9.2}{Rule~9.2}
{R9.3}{Rule~9.3}
{R9.4}{Rule~9.4}
{R9.5}{Rule~9.5}
{R10.1}{Rule~10.1}
{R10.2}{Rule~10.2}
{R10.3}{Rule~10.3}
{R10.4}{Rule~10.4}
{R10.5}{\textit{Rule~10.5}}
{R10.6}{Rule~10.6}
{R10.7}{Rule~10.7}
{R10.8}{Rule~10.8}
{R11.1}{Rule~11.1}
{R11.2}{Rule~11.2}
{R11.3}{Rule~11.3}
{R11.4}{\textit{Rule~11.4}}
{R11.5}{\textit{Rule~11.5}}
{R11.6}{Rule~11.6}
{R11.7}{Rule~11.7}
{R11.8}{Rule~11.8}
{R11.9}{Rule~11.9}
{R12.1}{\textit{Rule~12.1}}
{R12.2}{Rule~12.2}
{R12.3}{\textit{Rule~12.3}}
{R12.4}{\textit{Rule~12.4}}
{R12.5}{\textbf{Rule~12.5}}
{R13.1}{Rule~13.1}
{R13.2}{Rule~13.2}
{R13.3}{\textit{Rule~13.3}}
{R13.4}{\textit{Rule~13.4}}
{R13.5}{Rule~13.5}
{R13.6}{\textbf{Rule~13.6}}
{R14.1}{Rule~14.1}
{R14.2}{Rule~14.2}
{R14.3}{Rule~14.3}
{R14.4}{Rule~14.4}
{R15.1}{\textit{Rule~15.1}}
{R15.2}{Rule~15.2}
{R15.3}{Rule~15.3}
{R15.4}{\textit{Rule~15.4}}
{R15.5}{\textit{Rule~15.5}}
{R15.6}{Rule~15.6}
{R15.7}{Rule~15.7}
{R16.1}{Rule~16.1}
{R16.2}{Rule~16.2}
{R16.3}{Rule~16.3}
{R16.4}{Rule~16.4}
{R16.5}{Rule~16.5}
{R16.6}{Rule~16.6}
{R16.7}{Rule~16.7}
{R17.1}{Rule~17.1}
{R17.2}{Rule~17.2}
{R17.3}{\textbf{Rule~17.3}}
{R17.4}{\textbf{Rule~17.4}}
{R17.5}{\textit{Rule~17.5}}
{R17.6}{\textbf{Rule~17.6}}
{R17.7}{Rule~17.7}
{R17.8}{\textit{Rule~17.8}}
{R18.1}{Rule~18.1}
{R18.2}{Rule~18.2}
{R18.3}{Rule~18.3}
{R18.4}{\textit{Rule~18.4}}
{R18.5}{\textit{Rule~18.5}}
{R18.6}{Rule~18.6}
{R18.7}{Rule~18.7}
{R18.8}{Rule~18.8}
{R19.1}{\textbf{Rule~19.1}}
{R19.2}{\textit{Rule~19.2}}
{R20.1}{\textit{Rule~20.1}}
{R20.2}{Rule~20.2}
{R20.3}{Rule~20.3}
{R20.4}{Rule~20.4}
{R20.5}{\textit{Rule~20.5}}
{R20.6}{Rule~20.6}
{R20.7}{Rule~20.7}
{R20.8}{Rule~20.8}
{R20.9}{Rule~20.9}
{R20.10}{\textit{Rule~20.10}}
{R20.11}{Rule~20.11}
{R20.12}{Rule~20.12}
{R20.13}{Rule~20.13}
{R20.14}{Rule~20.14}
{R21.1}{Rule~21.1}
{R21.2}{Rule~21.2}
{R21.3}{Rule~21.3}
{R21.4}{Rule~21.4}
{R21.5}{Rule~21.5}
{R21.6}{Rule~21.6}
{R21.7}{Rule~21.7}
{R21.8}{Rule~21.8}
{R21.9}{Rule~21.9}
{R21.10}{Rule~21.10}
{R21.11}{Rule~21.11}
{R21.12}{\textit{Rule~21.12}}
{R21.13}{\textbf{Rule~21.13}}
{R21.14}{Rule~21.14}
{R21.15}{Rule~21.15}
{R21.16}{Rule~21.16}
{R21.17}{\textbf{Rule~21.17}}
{R21.18}{\textbf{Rule~21.18}}
{R21.19}{\textbf{Rule~21.19}}
{R21.20}{\textbf{Rule~21.20}}
{R21.21}{Rule~21.21}
{R22.1}{Rule~22.1}
{R22.2}{\textbf{Rule~22.2}}
{R22.3}{Rule~22.3}
{R22.4}{\textbf{Rule~22.4}}
{R22.5}{\textbf{Rule~22.5}}
{R22.6}{\textbf{Rule~22.6}}
{R22.7}{Rule~22.7}
{R22.8}{Rule~22.8}
{R22.9}{Rule~22.9}
{R22.10}{Rule~22.10}}%
[unknown MC3R1 Id #1]%
}

\newcommand*{\MCHd}[1]{%
\IfStrEqCase{#1}{%
{D2.1}{All source files shall compile without any compilation errors}
{D4.1}{Run-time failures shall be minimized}
{D4.3}{Assembly language shall be encapsulated and isolated}
{D4.8}{If a pointer to a structure or union is never dereferenced within a translation unit, then the implementation of the object should be hidden}
{D4.10}{Precautions shall be taken in order to prevent the contents of a \textit{header file} being included more than once}
{D4.12}{Dynamic memory allocation shall not be used}
{D4.14}{The validity of values received from external sources shall be checked}
{R1.1}{The program shall contain no violations of the standard C syntax and \textit{constraints}, and shall not exceed the implementation's translation limits}
{R1.2}{Language extensions should not be used}
{R1.4}{Emergent language features shall not be used}
{R6.1}{Bit-fields shall only be declared with an appropriate type}
{R7.4}{A string literal shall not be \textit{assigned} to an object unless the object's type is ``pointer to \code{const}-qualified \code{char}''}
{R8.6}{An identifier with external linkage shall have exactly one external definition}
{R9.1}{The value of an object with automatic storage duration shall not be read before it has been set}
{R12.5}{The \code{sizeof} operator shall not have an operand which is a function parameter declared as ``array of type''}
{R14.4}{The controlling expression of an \code{if} statement and the controlling expression of an \textit{iteration-statement} shall have \textit{essentially Boolean} type}
{R16.1}{All \code{switch} statements shall be well-formed}
{R16.4}{Every \code{switch} statement shall have a \code{default} label}
{R17.2}{Functions shall not call themselves, either directly or indirectly}
{R21.1}{\code{\#define} and \code{\#undef} shall not be used on a reserved identifier or reserved macro name}
{R21.2}{A reserved identifier or reserved macro name shall not be declared}
{R21.3}{The memory allocation and deallocation functions of \code{<stdlib.h>} shall not be used}
{R21.7}{The Standard Library functions \code{atof}, \code{atoi}, \code{atol} and \code{atoll} of \code{<stdlib.h>} shall not be used}
{R21.11}{The standard \textit{header file} \code{<tgmath.h>} shall not be used}
{R22.1}{All resources obtained dynamically by means of Standard Library functions shall be explicitly released}}%
[unknown MC3R1 Hd #1]%
}

\usepackage{afterpage}
\usepackage{enumitem}
\usepackage{flushend}
\usepackage{amssymb}
\usepackage{url}
\usepackage{graphicx}
\usepackage{fancyvrb}
\usepackage{times}
\usepackage[hidelinks]{hyperref}
\usepackage{xcolor}
\hypersetup{
    colorlinks,
    linkcolor={red!50!black},
    citecolor={blue!50!black},
    urlcolor={blue!80!black}
}

\newcommand*{\LTLM}{\hyperref[set:LTLM]{LTLM}}
\newcommand*{\DEVM}{\hyperref[set:DEVM]{DEVM}}
\newcommand*{\CTIS}{\hyperref[set:CTIS]{CTIS}}
\newcommand*{\RTIS}{\hyperref[set:RTIS]{RTIS}}
\newcommand*{\DOCU}{\hyperref[set:DOCU]{DOCU}}
\newcommand*{\ENMO}{\hyperref[set:ENMO]{ENMO}}
\newcommand*{\HTDR}{\hyperref[set:HTDR]{HTDR}}
\newcommand*{\EAPI}{\hyperref[set:EAPI]{EAPI}}
\newcommand*{\BAPI}{\hyperref[set:BAPI]{BAPI}}
\newcommand*{\PORT}{\hyperref[set:PORT]{PORT}}
\newcommand*{\CTIL}{\hyperref[set:CTIL]{CTIL}}
\newcommand*{\RTIL}{\hyperref[set:RTIL]{RTIL}}
\newcommand*{\TNTI}{\hyperref[set:TNTI]{TNTI}}
\newcommand*{\IRRC}{\hyperref[set:IRRC]{IRRC}}
\newcommand*{\TYPM}{\hyperref[set:TYPM]{TYPM}}
\newcommand*{\CSTR}{\hyperref[set:CSTR]{CSTR}}

\newcommand*{\coanote}[1]{}

\newenvironment{STDdescription}
               {\list{}{\labelwidth 0pt \itemindent-\leftmargin
                        \let\makelabel\descriptionlabel}}
               {\endlist}
\renewcommand*\descriptionlabel[1]{\hspace\labelsep
                                   \normalfont\bfseries #1}

\makeatletter
\newenvironment{btHighlight}[1][]
{\begingroup\tikzset{bt@Highlight@par/.style={#1}}\begin{lrbox}{\@tempboxa}}
{\end{lrbox}\bt@HL@box[bt@Highlight@par]{\@tempboxa}\endgroup}

\newcommand\btHL[1][]{%
  \begin{btHighlight}[#1]\bgroup\aftergroup\bt@HL@endenv%
}
\def\bt@HL@endenv{%
  \end{btHighlight}%
  \egroup
}
\newcommand{\bt@HL@box}[2][]{%
  \tikz[#1]{%
    \pgfpathrectangle{\pgfpoint{1pt}{0pt}}{\pgfpoint{\wd #2}{\ht #2}}%
    \pgfusepath{use as bounding box}%
    \node[anchor=base west, outer sep=0pt,inner xsep=1pt, inner ysep=0pt, rounded corners=2pt, minimum height=\ht\strutbox+1pt,#1]{\raisebox{1pt}{\strut}\strut\usebox{#2}};
  }%
}
\makeatother

\VerbatimFootnotes
\IEEEoverridecommandlockouts
\begin{document}

\title{A Rationale-Based Classification \\ of MISRA C Guidelines}
\author{\IEEEauthorblockN{Roberto Bagnara\textsuperscript{*}\thanks{%
{}\textsuperscript{*} Roberto Bagnara
is a member of the \emph{MISRA C Working Group} and of
ISO/IEC JTC1/SC22/WG14, a.k.a.\ the \emph{C Standardization Working Group}.
Patricia M.\ Hill is a member of the \emph{MISRA C++ Working Group}.
Nonetheless, the views expressed in this paper are the authors' and should
not be taken to represent the views of the mentioned working groups
and organizations.}}
\IEEEauthorblockA{University of Parma\\
Parma, Italy\\
Email: roberto.bagnara@unipr.it}
\and
\IEEEauthorblockN{Abramo Bagnara}
\IEEEauthorblockA{BUGSENG\\
Parma, Italy\\
Email: abramo.bagnara@bugseng.com}
\and
\IEEEauthorblockN{Patricia M.\ Hill\textsuperscript{*}}
\IEEEauthorblockA{BUGSENG\\
Parma, Italy\\
Email: patricia.hill@bugseng.com}}

\maketitle

\begin{abstract}
MISRA~C is the most authoritative language subset for the
C programming language that is a de facto standard
in several industry sectors where safety and security are of paramount
importance.
While MISRA~C is currently encoded in 175 guidelines
(coding rules and directives),
it does not coincide with them:
proper adoption of MISRA~C requires embracing
its preventive approach (as opposed to the ``bug finding'' approach)
and a documented development process where
justifiable non-compliances are authorized and recorded
as \emph{deviations}.
MISRA~C guidelines are classified along several axes in the
official MISRA documents.
In this paper, we add to these an orthogonal classification
that associates guidelines with their main rationale.
The advantages of this new classification are illustrated
for different kinds of projects, including those not (yet)
having MISRA compliance among their objectives.
\end{abstract}

\section{Introduction}
\label{sec:introduction}

The C programming language is the most used language for the development
of embedded systems, including those implementing critical functionality
of various kinds.  There are strong economic reasons for this success:
the possibility of writing concise code,
high efficiency,
easy access to hardware features,
ISO standardization,
the availability of tools such as C compilers,
a long history of usage.

Unfortunately, the same reasons that are behind C's strong points
are also the source of serious weaknesses:
hundreds of behaviors are not fully defined by the language,
which is also easy to misunderstand and open to all sorts of
abuses \cite{BagnaraBH18,BagnaraBH19}.

The solution adopted by industry to mitigate this problem
is very pragmatic: \emph{language subsetting}.
Namely, the most important functional safety
standards%
\footnote{IEC 61508 \cite{IEC-61508:2010} (industrial, generic),
ISO~26262 \cite{ISO-26262:2018} (automotive),
CENELEC~EN~50128 \cite{CENELEC-EN-50128:2011} (railways),
RTCA DO-178C \cite{RTCA-DO-178C} (aerospace) and
FDA's \emph{General Principles of Software Validation} \cite{FDA02}
(medical devices).}
either mandate or strongly recommend that critical applications
programmed in C only use a restricted subset of the language
such that the potential of committing possibly
dangerous mistakes is reduced.

MISRA~C is the most authoritative and most widespread subset for the
C programming language.
It is defined by a set of \emph{guidelines} that enable the use of C
for the development of safety- and/or security-related software
as well as the development of applications with
high integrity or high reliability requirements \cite{MISRA-C-2012-Revision-1}.

In this paper, we propose a classification of MISRA~C guidelines
that is complementary to the orthogonal classifications provided
in the official MISRA~C documents.
This new classification encodes the \emph{rationale} of the guideline,
that is, the reason why the guideline is in MISRA~C.
The ability of quickly conveying the nature of the guideline rationale
has several advantages:
\begin{itemize}
\item
  it reduces rule misunderstanding for those that do not take the time
  to read the full guideline specification;
\item
  it helps better decision-making in the choice between compliance
  and deviation;
\item
  it facilitates guideline prioritization for projects
  where MISRA compliance requirements arrived late in the development
  process;
\item
  it facilitates guideline selection for projects
  that do no (yet) have MISRA compliance requirements.
\end{itemize}

The plan of the paper is as follows:
Section~\ref{sec:misra-c-and-misra-compliance} introduces MISRA~C
and the notion of MISRA compliance;
Section~\ref{sec:a-new-rationale-based-classification-of-misra-c-guidelines}
defines the classification;
Section~\ref{sec:using-the-new-classification-for-compliance} discusses
the uses of the new classification in projects seeking MISRA compliance;
Section~\ref{sec:subsetting-misra-c-guidelines} touches the subject of
subsetting MISRA~C guidelines;
Section~\ref{sec:conclusion} concludes.

\section{MISRA~C and MISRA Compliance}
\label{sec:misra-c-and-misra-compliance}

The first edition of MISRA~C was published in 1998 \cite{MISRA-C-1998}
targeting the need ---emerged in the automotive industry
and reflected in 1994's MISRA
\emph{Development guidelines for vehicle based software}
\cite{MISRA-1994}---
for ``a restricted subset of a standardized structured language.''

The first edition of MISRA~C was very successful and it was adopted
also outside the automotive sector.
The wider industry use and the reported user experiences prompted a
major reworking of MISRA~C that resulted, in 2004, in the publication
of its second edition.

\begin{figure*}[p]
\centering
\includegraphics[width=\textwidth, height=0.9\textheight,keepaspectratio]{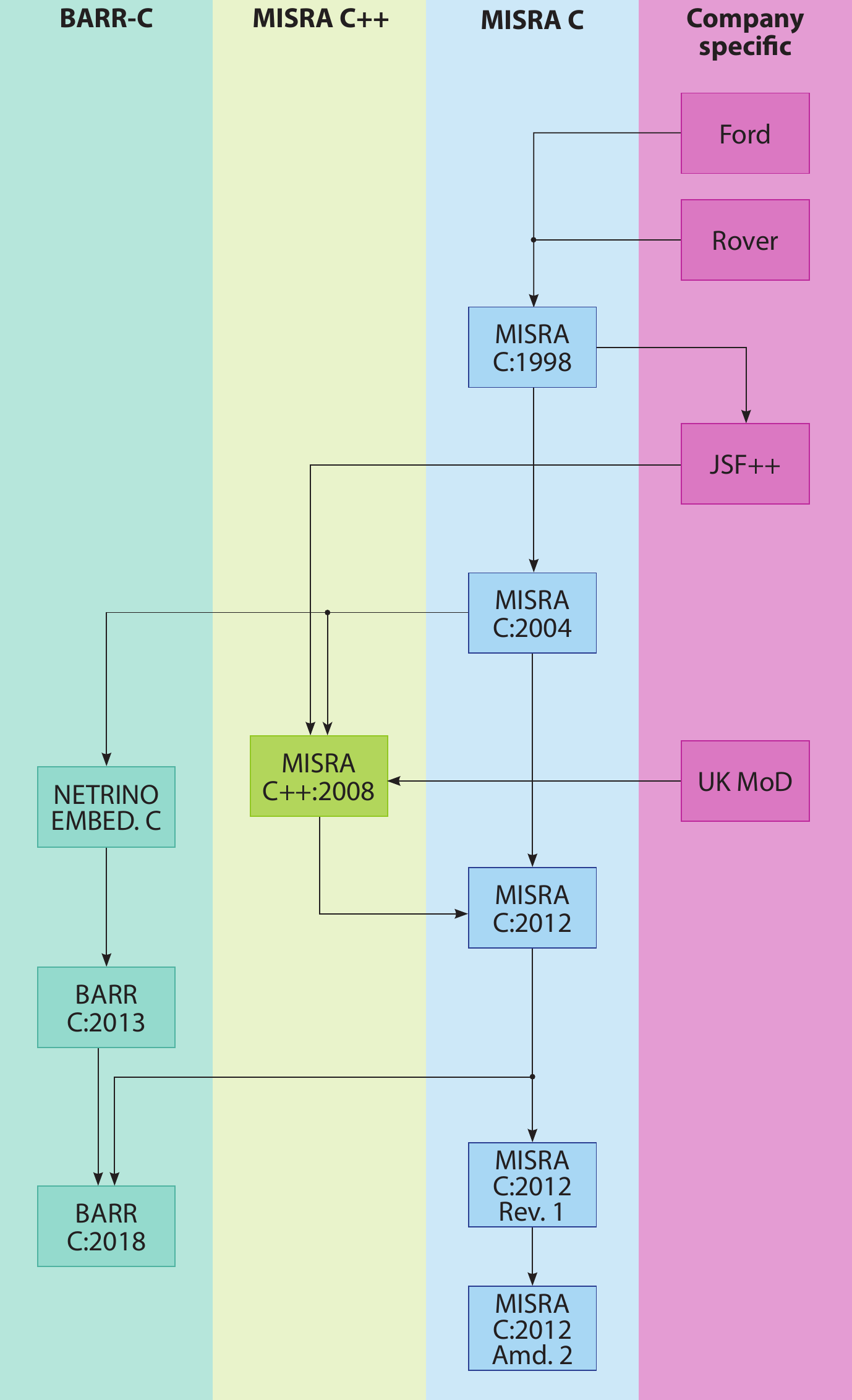}
\caption{Origin and history of MISRA C and BARR-C}
\label{fig:misra-c-history}
\end{figure*}

In 2013, after 4 years of renewed effort, the third edition of MISRA~C,
MISRA~C:2012, was published.
MISRA~C:2012 brought numerous improvements over the previous edition:
coverage of language issues was extended;
support for C99 \cite{ISO-C-1999-consolidated-TC3} was added in addition
to C90 \cite{ISO-C-1995};
mitigation for the lack of strong typing and corresponding type checking in C
was considerably improved.
In addition, the support for compliance was enhanced,
the guidelines were specified more clearly and precisely,
and, consequently, the likelihood that different static analysis tools
would give the same results was significantly increased.
Finally, MISRA~C:2012 provides expanded rationales and the guidelines
are more precisely targeted at real issues.

After the initial release of MISRA C:2012, the coding standard was
further amended with the addition of 14 new guidelines
\cite{MISRA-C-2012-Amendment-1} aimed at completing coverage
against ISO/IEC TS~17961:2013 ``C Secure''
\cite{ISO-IEC_DTS_17961-2013,MISRA-C-2012-Addendum-2-2}
and increasing coverage against CERT~C 2016 Edition
\cite{CERT-C-2016,MISRA-C-2012-Addendum-3}.
These amendments, along with \emph{MISRA C:2012 Technical Corrigendum 1}
\cite{MISRA-C-2012-TC1} have been consolidated into the first revision
of MISRA~C:2012 \cite{MISRA-C-2012-Revision-1}.
A further amendment \cite{MISRA-C-2012-Amendment-2}
extended the applicability of MISRA~C to programs written according to
the 2011 and 2018 standardized versions of the C language:
in the sequel they are simply denoted by \emph{C11} \cite{ISO-C-2011}
and \emph{C18} \cite{ISO-C-2018}.

MISRA~C, in its various versions, influenced all publicly-available
coding standards for C and C++ that were developed after MISRA~C:1998.
Figure~\ref{fig:misra-c-history} shows part of the relationship
and influence between the MISRA C/C++ guidelines and other sets of
guidelines.  The figure highlights the dependencies among MISRA~C editions,
revisions and amendments, and:
\begin{itemize}
\item
  Ford's and Land Rover's proprietary guidelines for the development of
  vehicle-based C software: these were merged into MISRA~C:1998;
 \item
   Lockheed Martin's \emph{JSF Air Vehicle C++ Coding Standards for
   the System Development and Demonstration Program} \cite{JSF-CPP-2005};
 \item
   BARR-C:2018 \cite{BARR-C-2018}
   and its ancestors: BARR-C:2013 \cite{BARR-C-2013} and the
   \emph{Netrino's Embedded~C Coding Standard} \cite{Netrino-C-2009};
 \item
   MISRA C++:2008, the C++ counterpart to MISRA~C, currently under revision
   \cite{MISRA-CPP-2008};
 \item
   the list of \emph{C++ vulnerabilities} that
   the UK Ministry of Defence contributed to MISRA~C++:2008
   \cite[Appendix~B]{MISRA-CPP-2008}.\footnote{This
     is the equivalent of Annex~J listing the various behaviors in ISO~C,
     which is missing in ISO~C++, making it hard work to identify them
     and to ensure they are covered by the guidelines.}
\end{itemize}
Many other items and relations are missing from the figure:
MISRA~C deeply influenced NASA's
``JPL Institutional Coding Standard for the C Programming Language''
\cite{NASA-JPL-C-2009} and several other coding standards (see, e.g.,
\cite{CERT-C-2016,IPA-ESCR-C-2014}).

In the sequel \emph{MISRA C} will denote \emph{MISRA~C:2012 Revision 1}
\cite{MISRA-C-2012-Revision-1} with \emph{Amendment~2}
\cite{MISRA-C-2012-Amendment-2}.

Each of the 175 guidelines of MISRA~C is classified as being either
a \emph{directive} or a \emph{rule}:
\begin{description}
\item[Rule:]
a guideline such that information concerning compliance
is fully contained in the source code \emph{and}
in the language implementation.
\item[Directive:]
a guideline such that information concerning compliance
is not fully contained in the source code and language implementation:
requirements, specifications, designs and other considerations
may need to be taken into account.
\end{description}

One of the things that is often misunderstood is that MISRA~C
is much more than just the set of its guidelines.
The guidelines are meant to be used in the framework of
a documented software development process.
The official document detailing what must be covered when making a claim
of MISRA compliance is MISRA~Compliance:2020 \cite{MISRA-Compliance-2020},
which also places constraints on such a development process.

The \emph{deviation process} is an essential part
of the adoption of the MISRA Guidelines, each one of which
is assigned a single category: \emph{mandatory},
\emph{required} or \emph{advisory}, defined as follows.
\begin{description}
\item[Mandatory:]
C code that complies with MISRA~C must comply with every
mandatory guideline: deviation is not permitted.
\item[Required:]
C code that complies with MISRA~C shall comply with every
required guideline: a formal deviation is required
where this is not the case.
\item[Advisory:]
these are recommendations that should be followed as far as it is
reasonably practical:
formal deviation is not required, but non-compliances should be
documented.\footnote{Advisory guidelines can also be \emph{disapplied}
  upon certain conditions, but we prefer not to dwell on these distinctions
  here.  The interested reader is referred to
  MISRA~Compliance:2020 \cite{MISRA-Compliance-2020} for all the details.}
\end{description}
Whenever complying with a guideline goes against code quality
or does not allow access to the hardware or does not allow
integrating or use suitably qualified adopted code, the guideline
has to be \emph{deviated}.  That is, instead of modifying the code to bring
it into compliance, for \emph{required} guidelines, a written argument
has to be provided to justify the violation whereas, for \emph{advisory}
guidelines, this is not necessary.

\section{A New, Rationale-Based Classification of MISRA~C Guidelines}
\label{sec:a-new-rationale-based-classification-of-misra-c-guidelines}

MISRA~C guidelines are classified along different axes
\cite{MISRA-C-2012-Revision-1}:
\begin{enumerate}
  \item The series to which the guidelines belong, that is, the first number
    in the numerical part of the guideline identifier.  For instance,
    Rule 8.1 belongs to rule series 8.  Series are in one-to-one correspondence
    with the \emph{main topic} of the guideline, and each one of them
    has a corresponding section in MISRA~C, where the section title is
    remindful of the topic.  For instance, rule series 7, 8 and 9
    are contained in the following sections of \cite{MISRA-C-2012-Revision-1}:
    \begin{description}
      \item[8.7] Literals and constants
      \item[8.8] Declarations and definitions
      \item[8.9] Initialization
    \end{description}
  \item The guideline being a \emph{rule} or a \emph{directive}.
    This is indicated by the guideline name, which always begins with `Rule' or
    `Dir' respectively.
  \item The guideline category: \emph{mandatory}, \emph{required} or
    \emph{advisory}. In this paper, identifiers for mandatory
    MISRA~C:2012 guidelines are set in boldface (e.g., \MCId{R9.1})
    whereas identifiers for advisory guidelines are set in italics
    (e.g., \MCId{D4.6}).
  \item The guideline being a \emph{decidable} or an \emph{undecidable}
    according to whether answering the question about compliance
    can be done algorithmically.
  \item The guideline scope being \emph{single translation unit} or
    \emph{system} according to the amount of code that needs to be
    analyzed in order to check compliance.
\end{enumerate}
All these classifications are defined in
MISRA~C.
While the last four axes have profound consequences on the notion
of \emph{MISRA compliance} and on the activities that tools and humans
may or may not have to perform, the first axis is just for
presentation convenience.  Moreover, there are guidelines that would
legitimately belong to more than one topic, and only the main one is
captured in the series.\footnote{Note that with our choice of fonts for
    the category, every guideline name presented here will indicate
    its classifications with respect to the first three axes.}

Here we propose a further classification of MISRA guidelines.
This classification encodes the main \emph{rationale} for each guideline,
that is, the nature of the addressed language and programming
issues.\footnote{In
  MISRA~C:2012 this is explained in a section titled ``Rationale''
  for each guideline \cite{MISRA-C-2012-Revision-1,MISRA-C-2012-Amendment-2}.}
According to the new classification each guideline belongs to one or two
of 16 named sets.
Only three guidelines belong to two sets, that is,
\MCId{R6.1}, \MCId{R8.6} and \MCId{R22.5}.

A synopsis of the new classification is presented in Figure~\ref{fig:synopsis}.
The following sections illustrate each named set.

\subsection{LTLM: Language/Toolchain/Library Misuse}
\label{set:LTLM}

This set contains 41 guidelines:
\MCId{D4.11},
\MCId{D4.13},
\MCId{R1.1},
\MCId{R1.3},
\MCId{R5.1},
\MCId{R5.2},
\MCId{R5.4},
\MCId{R6.1},
\MCId{R8.6},
\MCId{R8.10},
\MCId{R9.1},
\MCId{R9.4},
\MCId{R13.1},
\MCId{R13.2},
\MCId{R17.4},
\MCId{R17.5},
\MCId{R18.1},
\MCId{R18.2},
\MCId{R18.3},
\MCId{R18.6},
\MCId{R19.1},
\MCId{R20.2},
\MCId{R20.3},
\MCId{R20.4},
\MCId{R20.6},
\MCId{R20.11},
\MCId{R20.13},
\MCId{R20.14},
\MCId{R21.1},
\MCId{R21.2},
\MCId{R21.13},
\MCId{R21.14},
\MCId{R21.17},
\MCId{R21.18},
\MCId{R21.19},
\MCId{R21.20},
\MCId{R22.2},
\MCId{R22.4},
\MCId{R22.6},
\MCId{R22.8},
\MCId{R22.10}.

These guidelines are extremely important: they have to do with
the ``contract'' between those programming and reasoning about
programs in C\footnote{As opposed to programming and reasoning
  about programs in assembly language.} and the artifacts that
make this possible.  They are:
\begin{itemize}
\item
  The language specification: if the program violates the C syntax
  and constraints, or if it has undefined\footnote{This is ``behavior,
    upon use of a nonportable or erroneous program construct
    or of erroneous data, for which [the C standard]
    this document imposes no requirements'' \cite{ISO-C-2018}.}
  or critical unspecified behavior,\footnote{Unspecified behavior
    is ``behavior, that results from the use of an unspecified value,
    or other behavior upon which this document provides two or more
    possibilities and imposes no further requirements on which is
    chosen in any instance''  \cite{ISO-C-2018}. In MISRA~C an unspecified
    behavior is called \emph{critical} ``according to whether reliance
    on this behaviour is likely to lead to unexpected program operation''
    \cite[Appendix H.2]{MISRA-C-2012-Revision-1}.}
  then the program does not have a defined meaning.
\item
  The toolchain specification: if the program exceeds one or more
  of the translation limits\footnote{These are minimal quantities
    that conforming implementations have to meet or exceed.
    For instance, C90 mandates a minimum of at least 8 nesting levels
    of \code{#include}d files, 6 significant initial characters
    in an external identifier, 257 \code{case} labels for
    a \code{switch} statement \cite[Subclause 5.2.4.1]{ISO-C-1995}.
    While more recent versions of the ISO~C standard have more generous
    limits and many C implementations go beyond those limits,
    it must be noted that an implementation needs not issue a diagnostic
    message when a translation limit is exceeded.}
  of the language implementation (which, usually,
  is the combination of compiler, assembler, linker and librarian) then,
  again, the program does not have a defined meaning.
\item
  The library specification: if the program violates the preconditions
  of library functions and macros, or it fiddles with reserved identifiers,
  then, once more, the program does not have a defined meaning.
\end{itemize}

A very important guideline in this set is
\begin{quote}
\MCId{R1.1}: \MCHd{R1.1}.
\end{quote}
While this rule is not mandatory, it is one of the
cornerstones of MISRA C: if the program does not satisfy
the C language specification (modulo allowed extensions),
nothing can really be said about the program, let alone about
its MISRA compliance.
Even when MISRA compliance is not among the objectives,
violations point at important circumstances that otherwise
would completely escape the developers, such as:
\begin{itemize}
\item
  The mismatch between the language standard
  and the features used in the program.  For instance, the use of
  a C11-specific feature like \verb+_Static_assert+ in
  code compiled in C99 mode: perhaps the code should have been compiled
  in C11 or C18 mode, or perhaps \verb+_Static_assert+ was not meant
  to be used.
\item
  The use of documented language extensions, such as GCC's statement
  expressions.\footnote{See \url{https://gcc.gnu.org/onlinedocs/gcc/Statement-Exprs.html}.}
  Once they are accepted as extensions, they will fall under the scope of
  \MCId{R1.2}, which is covered later in this paper.
\item
  The use of constructs that violate the C standard syntax or constraints,
  for which the toolchain does not issue a diagnostic message and does not
  document as a supported extension.  This is the case, e.g., for the
  return of void expressions, undocumented but accepted without warning
  by GCC up to and including version 11.2.
  Of course, the lack of documentation makes it hard to legitimately
  consider such constructs as language extensions.
\end{itemize}

Another important rule in the LTLM set is:
\begin{quote}
\MCId{R9.1}: \MCHd{R9.1}.
\end{quote}
Reading an uninitialized object residing on the stack is undefined behavior,
so anything can happen.
\afterpage{%
\onecolumn
\begin{landscape}
\begin{figure}
\small
  \begin{tabular}{cccc}
    \begin{tabular}[t]{|p{1.4cm}p{1.4cm}p{1.4cm}|}
      \hline
      \multicolumn{3}{|c|}{\CTIS{}} \\
      \hline
      \MCId{D2.1} & \MCId{D4.10} & \\
      \hline
    \end{tabular} &
    \begin{tabular}[t]{|p{1.4cm}p{1.4cm}p{1.4cm}|}
      \hline
      \multicolumn{3}{|c|}{\RTIS{}} \\
      \hline
      \MCId{D4.1} & \MCId{R22.1} & \\
      \hline
    \end{tabular} &
    \begin{tabular}[t]{|p{1.4cm}p{1.4cm}p{1.4cm}|}
      \hline
      \multicolumn{3}{|c|}{\EAPI{}} \\
      \hline
      \MCId{R1.4} & & \\
      \hline
    \end{tabular} &
    \begin{tabular}[t]{|p{1.4cm}p{1.4cm}p{1.4cm}|}
      \hline
      \multicolumn{3}{|c|}{\BAPI{}} \\
      \hline
      \MCId{R21.7} & \MCId{R21.11} & \\
      \hline
    \end{tabular} \\
    \\
    \begin{tabular}[t]{|p{1.4cm}p{1.4cm}p{1.4cm}|}
      \hline
      \multicolumn{3}{|c|}{\DOCU{}} \\
      \hline
      \MCId{D1.1} & \MCId{D3.1} & \MCId{D4.2} \\
      \hline
    \end{tabular} &
    \begin{tabular}[t]{|p{1.4cm}p{1.4cm}p{1.4cm}|}
      \hline
      \multicolumn{3}{|c|}{\TNTI{}} \\
      \hline
      \MCId{D4.14} & & \\
      \hline
    \end{tabular} \\
    \\
    \begin{tabular}[t]{|p{1.4cm}p{1.4cm}p{1.4cm}|}
      \hline
      \multicolumn{3}{|c|}{\ENMO{}} \\
      \hline
      \MCId{D4.3} & \MCId{D4.8} & \MCId{R8.7} \\
      \MCId{R8.9} & & \\
      \hline
    \end{tabular} &
    \begin{tabular}[t]{|p{1.4cm}p{1.4cm}p{1.4cm}|}
      \hline
      \multicolumn{3}{|c|}{\PORT{}} \\
      \hline
      \MCId{D4.6} & \MCId{R1.2} & \MCId{R6.1} \\
      \MCId{R22.5} & & \\
      \hline
    \end{tabular} &
    \begin{tabular}[t]{|p{1.4cm}p{1.4cm}p{1.4cm}|}
      \hline
      \multicolumn{3}{|c|}{\RTIL{}} \\
      \hline
      \MCId{D4.7} & \MCId{R15.7} & \MCId{R16.4} \\
      \MCId{R17.7} & \MCId{R22.7} & \MCId{R22.9} \\
      \hline
    \end{tabular} \\
    \\
    \begin{tabular}[t]{|p{1.4cm}p{1.4cm}p{1.4cm}|}
      \hline
      \multicolumn{3}{|c|}{\IRRC{}} \\
      \hline
      \MCId{R2.1} & \MCId{R2.2} & \MCId{R2.3} \\
      \MCId{R2.4} & \MCId{R2.5} & \MCId{R2.6} \\
      \MCId{R2.7} & \MCId{R8.6} & \\
      \hline
    \end{tabular} \\
    \\
    \begin{tabular}[t]{|p{1.4cm}p{1.4cm}p{1.4cm}|}
      \hline
      \multicolumn{3}{|c|}{\CSTR{}} \\
      \hline
      \MCId{R13.5} & \MCId{R14.2} & \MCId{R15.1} \\
      \MCId{R15.2} & \MCId{R15.3} & \MCId{R15.4} \\
      \MCId{R15.5} & \MCId{R16.1} & \MCId{R16.2} \\
      \MCId{R16.5} & \MCId{R16.6} & \MCId{R18.4} \\
      \hline
    \end{tabular} &
    \begin{tabular}[t]{|p{1.4cm}p{1.4cm}p{1.4cm}|}
      \hline
      \multicolumn{3}{|c|}{\TYPM{}} \\
      \hline
      \MCId{R10.1} & \MCId{R10.2} & \MCId{R10.3} \\
      \MCId{R10.4} & \MCId{R10.5} & \MCId{R10.6} \\
      \MCId{R10.7} & \MCId{R10.8} & \MCId{R12.2} \\
      \MCId{R14.4} & \MCId{R16.7} & \MCId{R21.15} \\
      \hline
    \end{tabular} \\
    \\
    \begin{tabular}[t]{|p{1.4cm}p{1.4cm}p{1.4cm}|}
      \hline
      \multicolumn{3}{|c|}{\CTIL{}} \\
      \hline
      \MCId{R7.4} & \MCId{R8.2} & \MCId{R8.4} \\
      \MCId{R8.13} & \MCId{R11.1} & \MCId{R11.2} \\
      \MCId{R11.3} & \MCId{R11.4} & \MCId{R11.5} \\
      \MCId{R11.6} & \MCId{R11.7} & \MCId{R11.8} \\
      \MCId{R11.9} & \MCId{R17.3} & \\
      \hline
    \end{tabular} &
    \begin{tabular}[t]{|p{1.4cm}p{1.4cm}p{1.4cm}|}
      \hline
      \multicolumn{3}{|c|}{\HTDR{}} \\
      \hline
      \MCId{D4.12} & \MCId{R8.14} & \MCId{R12.3} \\
      \MCId{R14.1} & \MCId{R16.3} & \MCId{R17.1} \\
      \MCId{R17.2} & \MCId{R17.6} & \MCId{R18.5} \\
      \MCId{R18.7} & \MCId{R18.8} & \MCId{R19.2} \\
      \MCId{R20.1} & \MCId{R20.5} & \MCId{R20.10} \\
      \MCId{R20.12} & \MCId{R21.3} & \MCId{R21.4} \\
      \MCId{R21.5} & \MCId{R21.6} & \MCId{R21.8} \\
      \MCId{R21.9} & \MCId{R21.10} & \MCId{R21.12} \\
      \MCId{R21.16} & \MCId{R21.21} & \MCId{R22.3} \\
      \MCId{R22.5} & & \\
      \hline
    \end{tabular} &
    \begin{tabular}[t]{|p{1.4cm}p{1.4cm}p{1.4cm}|}
      \hline
      \multicolumn{3}{|c|}{\DEVM{}} \\
      \hline
      \MCId{D4.4} & \MCId{D4.5} & \MCId{D4.9} \\
      \MCId{R3.1} & \MCId{R3.2} & \MCId{R4.1} \\
      \MCId{R4.2} & \MCId{R5.3} & \MCId{R5.5} \\
      \MCId{R5.6} & \MCId{R5.7} & \MCId{R5.8} \\
      \MCId{R5.9} & \MCId{R6.2} & \MCId{R7.1} \\
      \MCId{R7.2} & \MCId{R7.3} & \MCId{R8.1} \\
      \MCId{R8.3} & \MCId{R8.5} & \MCId{R8.8} \\
      \MCId{R8.11} & \MCId{R8.12} & \MCId{R9.2} \\
      \MCId{R9.3} & \MCId{R9.5} & \MCId{R12.1} \\
      \MCId{R12.4} & \MCId{R12.5} & \MCId{R13.3} \\
      \MCId{R13.4} & \MCId{R13.6} & \MCId{R14.3} \\
      \MCId{R15.6} & \MCId{R17.8} & \MCId{R20.7} \\
      \MCId{R20.8} & \MCId{R20.9} & \\
      \hline
    \end{tabular} &
    \begin{tabular}[t]{|p{1.4cm}p{1.4cm}p{1.4cm}|}
      \hline
      \multicolumn{3}{|c|}{\LTLM{}} \\
      \hline
      \MCId{D4.11} & \MCId{D4.13} & \MCId{R1.1} \\
      \MCId{R1.3} & \MCId{R5.1} & \MCId{R5.2} \\
      \MCId{R5.4} & \MCId{R6.1} & \MCId{R8.6} \\
      \MCId{R8.10} & \MCId{R9.1} & \MCId{R9.4} \\
      \MCId{R13.1} & \MCId{R13.2} & \MCId{R17.4} \\
      \MCId{R17.5} & \MCId{R18.1} & \MCId{R18.2} \\
      \MCId{R18.3} & \MCId{R18.6} & \MCId{R19.1} \\
      \MCId{R20.2} & \MCId{R20.3} & \MCId{R20.4} \\
      \MCId{R20.6} & \MCId{R20.11} & \MCId{R20.13} \\
      \MCId{R20.14} & \MCId{R21.1} & \MCId{R21.2} \\
      \MCId{R21.13} & \MCId{R21.14} & \MCId{R21.17} \\
      \MCId{R21.18} & \MCId{R21.19} & \MCId{R21.20} \\
      \MCId{R22.2} & \MCId{R22.4} & \MCId{R22.6} \\
      \MCId{R22.8} & \MCId{R22.10} & \\
      \hline
    \end{tabular}
  \end{tabular}
  \caption{Synopsis of the rationale-based classification of MISRA~C guidelines}
  \label{fig:synopsis}
\end{figure}
\end{landscape}
\twocolumn
}

\noindent
In the most typical case,
memory or register values that, due to the previous computation history,
happen to be there, will be read.\footnote{Beware, this is not the only case.
  E.g., compilers can exploit undefined behavior in ways that may
  completely baffle programmers.}
The production of unexpected program results is of course unacceptable
in critical applications, but there is something worse:
code violating this rule is vulnerable to exploits.
Exploits can be of two kinds: (1) confidential information is leaked
to unauthorized parties; (2) attackers are given ways to control the system.
For the first kind, consider the following example, which is a simplified
version of a real bug \cite{Bastien19}:
\begin{Ccode}
int get_hw_address(struct device *dev,
                   struct user *usr) {
  unsigned char addr[MAX_ADDR_LEN];
  if (!dev->has_address)
    return -EOPNOTSUPP;
  dev->get_hw_address(addr);
  return copy_out(usr, addr, sizeof(addr));
}
\end{Ccode}
Automatic array variable \code{addr} declared in line~3 may only be partly
initialized by the indirect function call at line~6, but in any case
the entire contents of the array is copied from kernel space to user
space and that may constitute an important brick to construct a more
complex, reliable exploit.
For the second kind of exploit, consider the following example \cite{Bastien19}:
\begin{Ccode}
int queue_manage() {
  struct async_request *backlog;

  if (engine->state == IDLE)
    backlog = get_backlog(&engine->queue);

  if (backlog)
    backlog->complete(backlog, -EINPROGRESS);

  return 0;
}
\end{Ccode}
Here, the automatic pointer variable \code{backlog} is uninitialized at the
point of declaration in line~2.  The guard of the \code{if} statement
at line~4 is usually true, so \code{backlog} is usually initialized
at line~5 and testing is unlikely to find the issue.
However, when the statement at line~5 is not executed, an attacker
who can control the value of \code{backlog} can cause the application
to call any function that the attacker stores at the address in
\code{backlog->complete}.

A common form of language/toolchain/library misuse is captured
by the following guidelines:
\begin{quote}
\MCId{R21.1}: \MCHd{R21.1}.
\end{quote}

\begin{quote}
\MCId{R21.2}: \MCHd{R21.2}.
\end{quote}
It is as if someone, in the distant past, thought that it was a good
idea to define macros and other identifiers beginning with a leading
underscore in user code.  Maybe that person looked at standard library
header files and concluded that doing so was good style, and this
false belief was somehow propagated and perpetuated.
Whatever the reason is, it is a matter of fact that projects that have not
previously undergone MISRA compliance checking tend to have hundreds or
thousands of violations of these guidelines.
This is a potential source of undefined behavior and, as such,
something to definitely avoid.

\subsection{DEVM: Developer Misreading/Mistyping}
\label{set:DEVM}

This set contains 38 guidelines:
\MCId{D4.4},
\MCId{D4.5},
\MCId{D4.9},
\MCId{R3.1},
\MCId{R3.2},
\MCId{R4.1},
\MCId{R4.2},
\MCId{R5.3},
\MCId{R5.5},
\MCId{R5.6},
\MCId{R5.7},
\MCId{R5.8},
\MCId{R5.9},
\MCId{R6.2},
\MCId{R7.1},
\MCId{R7.2},
\MCId{R7.3},
\MCId{R8.1},
\MCId{R8.3},
\MCId{R8.5},
\MCId{R8.8},
\MCId{R8.11},
\MCId{R8.12},
\MCId{R9.2},
\MCId{R9.3},
\MCId{R9.5},
\MCId{R12.1},
\MCId{R12.4},
\MCId{R12.5},
\MCId{R13.3},
\MCId{R13.4},
\MCId{R13.6},
\MCId{R14.3},
\MCId{R15.6},
\MCId{R17.8},
\MCId{R20.7},
\MCId{R20.8},
\MCId{R20.9}.

Guidelines in this set have to do with the prevention of mistakes
arising from three phenomena:
\begin{enumerate}
\item
  developers not being aware of all the C language intricacies;
\item
  developers misreading or misinterpreting program elements;
\item
  developers typing something different from what was intended.
\end{enumerate}
The first two phenomena are important (but not the only) sources
of what MISRA~C refers to as \emph{developer confusion}.

An example guideline in this set is
\begin{quote}
\MCId{R12.5}: \MCHd{R12.5}.
\end{quote}
It is likely that the developer is not aware that the actual parameter
type is a pointer instead of the written array.
This guideline prevents potentially catastrophic consequences
of this possibility.

\subsection{CTIS: Compile-Time Issues}
\label{set:CTIS}

This set contains 2 guidelines:
\MCId{D2.1},
\MCId{D4.10}.

These directives guard against issues that may occur at compile time
whereby executable code does not match the source code that was meant
to produce it.

The first directive is actually a constraint on the
interface between the toolchain and the build procedure:
\begin{quote}
\MCId{D2.1}: \MCHd{D2.1}.
\end{quote}
A build procedure that is unable to reliably detect compilation errors
may link old or invalid object code into the application executable.

The second directive prevents including a header file multiple times:
\begin{quote}
\MCId{D4.10}: \MCHd{D4.10}.
\end{quote}
The systematic use of guards against multiple inclusion
is required for three reasons:
\begin{enumerate}
\item To avoid circular inclusion chains, which may give rise to
  compilation errors that are difficult to diagnose and understand,
  as well as undefined behavior (compilers are not required to detect
  circular inclusion chains).
\item To avoid mistakes in the ordering of header file inclusions.
  Without guards, programmers have to use a strict inclusion discipline
  in order to avoid compilation errors.
  Such a discipline that has been popular in the past is ``no header file can
  include another header file.''
  However, this results in non-header source files beginning with
  long lists of inclusions \emph{that have to be specified in the right order};
  getting the order wrong might result in a compilation error
  or, what is more important, unexpected behavior.
\item To avoid unnecessary long compilation time and static analysis
  time, the latter with the danger that a complexity-throttling
  measure in the static analyzer might lead to analysis imprecision.
\end{enumerate}

\subsection{RTIS: Run-Time Issues}
\label{set:RTIS}

This set contains 2 guidelines:
\MCId{D4.1},
\MCId{R22.1}.

These guidelines cover run-time issues that are not covered by the
rules in other sets.
\begin{quote}
\MCId{D4.1}: \MCHd{D4.1}.
\end{quote}
This directive requires covering run-time failure by any set of measures:
prevention, testing, run-time detection, static analysis and more.
The adopted measures should of course be documented to be amenable to
peer review.

\begin{quote}
\MCId{R22.1}: \MCHd{R22.1}.
\end{quote}
Program resources (e.g., files, memory, synchronization devices, etc.)
are all in finite supply: failure to release them,
a.k.a.\ \emph{resource leakage}, will, sooner or later, lead to
their exhaustion and the consequent potential program misbehavior.

\subsection{DOCU: Documentation}
\label{set:DOCU}

This set contains 3 guidelines:
\MCId{D1.1},
\MCId{D3.1},
\MCId{D4.2}.

The documentation to be provided concerns:
\begin{itemize}
\item
implementation-defined behaviors affecting the program semantics
(\MCId{D1.1});
\item
traceability information to requirements
(\MCId{D3.1});
\item
assembly code
(\MCId{D4.2}).
\end{itemize}
Given the uncertainty about the requirements to be implemented,
about the implementation-defined behaviors of the selected toolchain,
and about the meaning of assembly code fragments, procrastinating
compliance with these guidelines is not advisable.

\subsection{ENMO: Encapsulation/Modularization}
\label{set:ENMO}

This set contains 4 guidelines:
\MCId{D4.3},
\MCId{D4.8},
\MCId{R8.7},
\MCId{R8.9}.

Encapsulation of programming procedures by functions and macros makes a program
easier to read and modify in the future.  Modularization divides a
function into an interface (a header file) and an implementation (the
source file). Users of the function can only have access to the
header file, and the header file contents should be minimized.

Consider the following directive:
\begin{quote}
\MCId{D4.3}: \MCHd{D4.3}.
\end{quote}
By isolating the assembly language in functions or macros, the
readability and maintainability of such code is improved. Moreover, this
encapsulation ensures that the code can more easily be later replaced by
alternative code with an equivalent implementation.

\begin{quote}
\MCId{D4.8}: \MCHd{D4.8}.
\end{quote}
When header files contain too much detail, chances are that such
detail is depended upon in unwanted ways.  This directive prescribes
the use of \emph{opaque pointers} to hide the implementation details
of data structures from user code that does not need to interfere with them.

\subsection{HTDR: Hard To Do Right}
\label{set:HTDR}

This set contains 28 guidelines:
\MCId{D4.12},
\MCId{R8.14},
\MCId{R12.3},
\MCId{R14.1},
\MCId{R16.3},
\MCId{R17.1},
\MCId{R17.2},
\MCId{R17.6},
\MCId{R18.5},
\MCId{R18.7},
\MCId{R18.8},
\MCId{R19.2},
\MCId{R20.1},
\MCId{R20.5},
\MCId{R20.10},
\MCId{R20.12},
\MCId{R21.3},
\MCId{R21.4},
\MCId{R21.5},
\MCId{R21.6},
\MCId{R21.8},
\MCId{R21.9},
\MCId{R21.10},
\MCId{R21.12},
\MCId{R21.16},
\MCId{R21.21},
\MCId{R22.3},
\MCId{R22.5}.

These guidelines warn against the use of harmful, but possibly very useful
language features that require considerable expertise if they have to be
used in a critical system.

As an example, let us consider the required directive
\begin{quote}
\MCId{D4.12}: \MCHd{D4.12}.
\end{quote}
There is an increasing demand to allow dynamic memory allocation,
but insufficient understanding about the consequences of using
this technique.
Since \MCId{D4.12} is a required directive, it can be deviated with
a proper motivation.
However, formulating such a motivation in a defensible way,
requires dealing with a number of issues, summarized below,
that are not easy to solve.
Note that such issues have to be dealt with anyway, even
when not following MISRA~C.

\subsubsection{Out-of-storage Run-Time Failures}

It is often not easy to ensure that the available heap memory will
always be sufficient for the chosen allocation strategy.  There are
various possibilities:

\begin{enumerate}
  \item[a.]
    Available memory is demonstrably sufficient even if allocated memory
    is never deallocated.  This is a very favorable situation:
    we can avoid releasing memory and, doing so,
    we need not be concerned with memory allocation errors.
  \item[b.]
    Available memory is not sufficient if we do not deallocate.  Then we
    will have to release memory when it is no longer used.  There
    are two options:
    \begin{enumerate}
    \item[b1.]
      Explicit release: then we are confronted with dangling
      pointers (when we release too early), memory leaks (when we
      release too late or fail to release), double-free errors (when
      we release more than once).  The use of smart pointers
      (preferably assisted by the use of automatic checkers to ensure
      they are used properly) can prevent all such errors;
      of course, using smart pointers comes with some space and time
      overhead that may or may not be appropriate depending on the
      application.
    \item[b2.]  Automatic release, i.e., via the use of a garbage
      collector: we avoid dangling pointers, memory leaks and
      double-free errors, but the execution time of our code becomes
      more difficult to predict (the so-called real-time, incremental
      garbage collectors do not completely solve the ``stop-the-world
      effect'').
    \end{enumerate}
\end{enumerate}

In both cases, proving that the space is sufficient is very often
difficult.  One of the reasons is fragmentation, which is a
particularly important problem for long-running systems (systems
that will run continuously for many hours, days or even weeks):
depending on the distribution of memory allocation sizes, the free
memory can be filled by very small fragments that will have to
be compacted (this is not always possible and may cause
``stop-the-world effect'')
or they might cause an out-of-storage condition.

\subsubsection{Timing Considerations}

Memory allocation and deallocation functions are often subject to
variable (and possibly long) latency: free lists or similar
data-structures have to be searched and maintained, free blocks have
to be searched, merged, splitted and possibly moved.
This is unacceptable for systems that have to meet hard real-time constraints.

In order to better understand what is the best strategy for a project,
one needs to answer the following questions (in no particular order):
\begin{itemize}
  \item
    Do we have hard real-time constraints?
  \item
    Do we have tight limits on the available computing power or can
    we trade some computing power in exchange for correctness guarantees?
  \item
    Is there a simple pattern in the memory sizes we need to dynamically
    allocate?  (There are relatively easy, partial solutions for the
    fragmentation problem if we only need blocks of, say, a dozen fixed
    dimensions.)
  \item
    Is our system long-running or can we assume it will be rebooted
    at least once in a given number of hours or days?
  \item
    Is there a demonstrable upper bound to the amount of memory
    that has to be dynamically allocated?  Is this upper bound
    known at system-startup?
\end{itemize}
The rationale of the last question is the following: if dynamic memory
allocation could take place at system-startup only, possibly depending
on some configuration parameters, then the check for memory
sufficiency is done during startup together with other startup
self-test activities.  This is the recommendation given by the
\emph{JSF Air Vehicle C++ Coding Standards for
the System Development and Demonstration Program} \cite{JSF-CPP-2005}
coding standard with the advantage that no dynamic memory
allocation errors can occur once the system is operating in a safety-critical
mode.

While \MCId{D4.12} is concerned with programming technique in general,
however it is implemented, another rule in the HTDR set targets
the typical implementations provided as part of the C standard library:
\begin{quote}
\MCId{R21.3}: \MCHd{R21.3}.
\end{quote}
In addition to the issues mentioned above, such implementations,
being general-purpose and not accompanied by solid specifications,
have unknown fragmentation and timing properties.

A final example of a rule cautioning against a program feature that
may be both exceedingly useful (for some applications) and exceedingly
complex to get right is the following:
\begin{quote}
\MCId{R17.2}: \MCHd{R17.2}.
\end{quote}
The possibility of writing (mutually) recursive functions greatly simplifies
the task of writing code operating on non-trivial data structures defined
by means of recursive rules (such as languages defined by context-free
grammars).  This, however, is not a frequent requirement for critical systems.
In contrast, the presence of (mutually) recursive functions considerably
complicates the task of bounding the maximum stack size needed to run the
code.  As is well known, stack overflow is a serious and potentially
catastrophic issue (see, e.g., \cite{Koopman14}).

\subsection{EAPI: Emergent APIs}
\label{set:EAPI}

This set contains just 1 guideline:
\MCId{R1.4}.

\begin{quote}
\MCId{R1.4}: \MCHd{R1.4}.
\end{quote}
This is the counterpart to \MCId{R1.1} that allows using
language implementations conforming to
C11 \cite{ISO-C-2011} and C18 \cite{ISO-C-2018},
except for the features that are not yet covered by MISRA~C,
namely:
\begin{itemize}
\item
  type generic expressions: \code{_Generic};
\item
  support for no-return functions: \code{_Noreturn}, \code{<stdnoreturn.h>};
\item
  support for multiple threads of execution:
  \code{_Atomic}, \code{<stdatomic.h>},
  \code{_Thread_local}, \code{<threads.h>};
\item
  support for the alignment of objects:
  \code{_Alignas}, \code{_Alignof}, \code{<stdalign.h>};
\item
  bounds-checking interfaces of C11/C18 Annex~K.
\end{itemize}

\subsection{BAPI: Badly-designed/Obsolete APIs}
\label{set:BAPI}

This set contains 2 guidelines:
\MCId{R21.7},
\MCId{R21.11}.

Standard library functions that have restricted input values have a
variety of ways to inform the caller that an argument's value is wrong.
These include returning a value such as \code{-1} or setting \code{errno}
to an error value.
However, this does not apply to all library functions.
Consider the rule
\begin{quote}
\MCId{R21.7}: \MCHd{R21.7}.
\end{quote}
These functions convert the initial portion of the string pointed
to by its parameter to an arithmetic type.
However, when the result cannot be represented, the behavior is undefined.
As these functions have been superseded by
\code{strtod}, \code{strtof}, \code{strtold}
\code{strtol} and \code{strtoll}, they can be considered both
badly designed and obsolete.

Obsolescence is not the reason behind rule
\begin{quote}
\MCId{R21.11}: \MCHd{R21.11}.
\end{quote}
Here the point is that the use of the type-generic features provided
by \code{tgmath.h} can have undefined behavior.

\subsection{PORT: Portability}
\label{set:PORT}

This set contains 4 guidelines:
\MCId{D4.6},
\MCId{R1.2},
\MCId{R6.1},
\MCId{R22.5}.

Guidelines in this set have portability as their main objective.
Consequently, for a project that is not meant to be portable and where
its developers have complete knowledge of all the
implementation-defined behaviors, this set may not seem to be very
relevant apart from one notable exception
Consider the rule
\begin{quote}
\MCId{R1.2}: \MCHd{R1.2}.
\end{quote}
A program that relies on extensions will be difficult to
port to a different language implementation,
and hence adherence to this rule will help in ensuring its portability.
But this is not the only reason why we have this rule.
In order to code and reason about programs in C we need to build
confidence on the correctness of the used C compiler.
Functional safety standards such as ISO~26262 \cite{ISO-26262:2018}
and RTCA DO-178C \cite{RTCA-DO-178C} prescribe
\emph{compiler qualification}.  This is usually achieved by running
\emph{validation suites} comprising tens of thousands of C test programs
through the compiler and checking the resulting outcome.
However, available C validation suites test the standardized language, not
the compiler extensions: this is another important reason why
the use of extensions should be limited as much as possible.

Another rule in this set is
\begin{quote}
\MCId{R6.1}: \MCHd{R6.1}.
\end{quote}
This covers three issues:
\begin{enumerate}
\item
  A bit-field type specified as unqualified \code{int}
  may be signed or unsigned depending on the implementation,
  hence, a portability issue.
\item
  Using bit-field types different from \code{unsigned int}
  or \code{signed int} is undefined behavior in C90:
  this is the reason why \MCId{R6.1} also occurs in the \LTLM{} set.
\item
  For later versions of ISO~C, a conforming compiler supports also
  \code{_Bool} as well as an implementation-defined set of integer types
  for bit-fields.  The used type may influence the implementation-defined
  aspects of bit-field layout.
\end{enumerate}

\subsection{CTIL: Compile-Time Information Loss}
\label{set:CTIL}

This set contains 14 guidelines:
\MCId{R7.4},
\MCId{R8.2},
\MCId{R8.4},
\MCId{R8.13},
\MCId{R11.1},
\MCId{R11.2},
\MCId{R11.3},
\MCId{R11.4},
\MCId{R11.5},
\MCId{R11.6},
\MCId{R11.7},
\MCId{R11.8},
\MCId{R11.9},
\MCId{R17.3}.

The objectives of the rules in this set include making sure
the compiler is given all the information it needs to diagnose
common and dangerous programming mistakes.
A good example is
\begin{quote}
\MCId{R7.4}: \MCHd{R7.4}.
\end{quote}
This rule demands that the read-only nature of string literals
is propagated using the \verb+const+ type qualifier,
allowing the compiler to flag possible misuses.

\subsection{RTIL: Run-Time Information Loss}
\label{set:RTIL}

This set contains 6 guidelines:
\MCId{D4.7},
\MCId{R15.7},
\MCId{R16.4},
\MCId{R17.7},
\MCId{R22.7},
\MCId{R22.9}.

Inadvertently corrupting, ignoring or erasing a run-time result
is very dangerous.
\begin{quote}
\MCId{R16.4}: \MCHd{R16.4}.
\end{quote}
Compliance with this rule ensures that a \verb+switch+ handles
\emph{all} cases, including, e.g., the values of an enumerated type
not corresponding to any enumerator.

\subsection{TNTI: Tainted Input}
\label{set:TNTI}

This set contains just 1 guideline:
\MCId{D4.14}.

Any input from external sources may be invalid due to accidental error
or malicious intent.
For instance:
\begin{itemize}
\item
  A value used to determine an array index could cause an array bounds error,
\item
  A value used to control a loop could cause (almost) infinite iterations,
\item
  A value used to compute a divisor may make it zero,
\item
  A value used to determine an amount of dynamic memory may lead
  to excessive memory allocation,
\item
  A string used as a query to an SQL database may contain a `\code{;}'
  character.
\end{itemize}
The rule
\begin{quote}
\MCId{D4.14}: \MCHd{D4.14}
\end{quote}
requires that all such input is checked appropriately.

\subsection{IRRC: Irrelevant Code}
\label{set:IRRC}

This set contains 6 guidelines:
\MCId{R2.1},
\MCId{R2.2},
\MCId{R2.3},
\MCId{R2.4},
\MCId{R2.5},
\MCId{R2.6},
\MCId{R2.7},
\MCId{R8.6}.

The coincidence of this set with series 2 of MISRA~C (\emph{Unused code})
is almost complete.  The exception is for rule
\begin{quote}
\MCId{R8.6}: \MCHd{R8.6}.
\end{quote}
for the cases where it does not give rise to undefined behavior.
It is for the other cases that this rule also occurs in the \LTLM{} set
of guidelines.

\subsection{TYPM: Types Misuse}
\label{set:TYPM}

This set contains 12 guidelines:
\MCId{R10.1},
\MCId{R10.2},
\MCId{R10.3},
\MCId{R10.4},
\MCId{R10.5},
\MCId{R10.6},
\MCId{R10.7},
\MCId{R10.8},
\MCId{R12.2},
\MCId{R14.4},
\MCId{R16.7},
\MCId{R21.15}.

The C language definition relies strongly on implicit type conversion.
Rather frequently, such conversions do not correspond to the will
of the programmer.  Sometimes this mismatch is due to a language
standard misinterpretation, sometimes it may be due to a typo.
Consider the following rule:
\begin{quote}
\MCId{R14.4}: \MCHd{R14.4}.
\end{quote}
This asks that the guards are effectively Boolean to ensure
the programmer intention is unambiguously expressed and matches the
true-or-false nature of controlling expressions.

\subsection{CSTR: Code Structuring}
\label{set:CSTR}

This set contains 12 guidelines:
\MCId{R13.5},
\MCId{R14.2},
\MCId{R15.1},
\MCId{R15.2},
\MCId{R15.3},
\MCId{R15.4},
\MCId{R15.5},
\MCId{R16.1},
\MCId{R16.2},
\MCId{R16.5},
\MCId{R16.6},
\MCId{R18.4}.

These guidelines define a subset of C that is well-structured so that
it is easier to review, maintain and analyze.
There are several rules that define a restricted format for the
selection and iteration statements.
As an example, consider the following rule:
\begin{quote}
\MCId{R16.1}: \MCHd{R16.1}.
\end{quote}
Depending on the value of its controlling expression, a \code{switch}
statement jumps to the code following a matching \code{case} label or, if there
is no match, to a \code{default} label if present, or, if not,
to the next statement.
Associated \code{break} statements will cause control to jump out of
the \code{switch} statement to the next statement.
However, the C syntax for the body of a \code{switch} is completely general
and these labels can be mixed with the code in the body
in an arbitrary way.

\MCId{R16.1} restricts the \code{switch} statement
to be a simple coherent structure where the body is a compound
statement and all associated \code{case} and \code{default} labels
are direct children of this block.  With consistent indentation of
dependent code, it should be straightforward to see which
\code{switch} the \code{case} and \code{default} labels are
associated to, and hence make it easier to examine how the different
cases are handled.

\section{Using the New Classification for MISRA Compliance}
\label{sec:using-the-new-classification-for-compliance}

The new classification of MISRA~C guidelines proposed in this paper,
being rationale-based, has the advantage of reminding programmers
of the guidelines' rationale.  This, in turn:
\begin{enumerate}
\item
  allows them to quickly and more faithfully reconstruct a guideline's
  objectives and, consequently,
\item
  improve their ability in formulating a defensible choice in the
  \emph{comply vs deviate} decision.
\end{enumerate}

While formal training is one of the requirements for MISRA compliance
\cite[Section 7.1]{MISRA-Compliance-2020}, programmers cannot generally
be expected to have the details of MISRA guidelines fresh in their minds.
Of course, for such programmers, the right thing to be done is to refresh
knowledge of the guideline and, in particular, of its rationale.
Unfortunately, not all programmers when in doubt will open the
relevant MISRA documents.

Independently of this, the decision, whether to comply with a MISRA
guideline or to deviate from it, is important.  Therefore a quick
reminder of the nature of the guideline rationale via our proposed new
classification is helpful to both programmers and project managers to
determine the subsequent steps.

Here is a non-exclusive set of questions for each rationale-based set.
When confronted with a violation of a guideline, the programmer can
check its rationale-based classification and, using the following list
(possibly enhanced with further pertinent queries), consider their
response to the associated questions:
\begin{STDdescription}
\item[\LTLM{}:] Do I really want to do that?  (Hint: probably not.)
\item[\DEVM{}:] Did I make a typing mistake?  Can I clarify my intentions?
\item[\CTIS{}:] Do I trust my build procedure?  Do I have multiple inclusion guards in place?
\item[\RTIS{}:] Did I exclude or mitigate possible run-time errors?
\item[\DOCU{}:] Is this documentation available? (Hint: if the answer is ``no'', then there is a problem.)
\item[\ENMO{}:] Will we need to maintain this program? (Hint: probably yes; we are not prescient.)
\item[\HTDR{}:] Did I consider the intricacies of this language feature?
\item[\EAPI{}:] Do I really want to use this API? (Hint: probably not.)
\item[\BAPI{}:] Do I really want to use this API? (Hint: definitely not.)
\item[\PORT{}:] Is portability among the project objectives?
\item[\CTIL{}:] Can I get this right without compiler assistance?
\item[\RTIL{}:] Am I losing valuable information here?
\item[\TNTI{}:] Is this input tainted?  If so, can it cause harm?
\item[\IRRC{}:] Why is this code here?
\item[\TYPM{}:] Can this result into a run-time type error?  Am I willing to run the risk?
\item[\CSTR{}:] Can I make this code easier to understand?
\end{STDdescription}

\section{Subsetting MISRA~C Guidelines}
\label{sec:subsetting-misra-c-guidelines}

Whereas the notion of MISRA compliance has degrees of flexibility,
to go beyond what is prescribed by MISRA~Compliance:2020
is non-negotiable.
In particular, in MISRA~C, there is no grading for the importance of
guidelines other than the one implied by the
\emph{mandatory}/\emph{required}/\emph{advisory} categorization.
That is, \emph{mandatory} guidelines are the most important ones,
then come \emph{required}, then come \emph{advisory} guidelines;
all required guidelines, whether rules or directives are of
equal importance, as are all mandatory guidelines,
as are all advisory ones.
A project cannot be MISRA compliant if some guidelines are
completely neglected: all have to be given appropriate consideration.

So it is clear that subsetting the MISRA~C Guidelines beyond what is allowed by
MISRA~Compliance:2020 does not allow MISRA~compliance of the project.
However, two things, expounded in the following sections, have to be
taken into account.

\subsection{No MISRA~Compliance Requirements for Most Projects}
\label{sec:no-misra-compliance-requirements-for-most-projects}

The vast majority of projects and organizations do not (yet) have
a MISRA~C requirement.
In addition, far too many projects in somewhat critical embedded systems,
are at the stage ``no coding standard, no static analysis''.
For these projects, in \cite{BagnaraBH21},
we proposed considering compliance with (subsets of) the BARR-C:2018
coding standard \cite{BARR-C-2018}.

Here we propose an alternative, namely, the (incremental)
adoption of subsets of MISRA~C.
The rational-based classification allows the simplified identification
of the guidelines that are more relevant to the project objectives.
For instance, if portability is among the objectives, the guidelines dealing
with portability are better included in the project coding standard.

One advantage of using subsets of MISRA~C is the availability
of good tools supporting it.  For projects leaning on preventive
measures, this is much better than resorting to bug finders, which
are notoriously plagued with false negatives and will thus only
discover parts of the issues.

A good strategy for a C project without MISRA compliance requirements
could be to enforce \LTLM{} guidelines first: they concern harmful
situations that should always be avoided one way or the other.  If
time allows, the \DEVM{} guidelines should be considered next: even
though violations of these guidelines are possibly harmless, they are
generally easily avoidable and solving them reduces the risk of
misleading others.  Then other rationale-based classification sets
can be considered depending on the context: use \PORT{} for portability,
\HTDR{} if inexperienced programmers are in the team, \ENMO{} if
maintainability and the possibility of reusing code for other projects
is contemplated.

\subsection{MISRA~Compliance Is Often Undertaken Late}
\label{sec:misra-compliance-is-often-undertaken-late}

Even for projects that do have MISRA~C requirements, it is a matter
of fact that the work on such requirements often begins
far too late in the development cycle.
It is well known that the highest payoff from the adoption of MISRA~C
is achieved when the MISRA guidelines are systematically enforced
from the start of the project or, at the very least,
before the code development reaches the review and unit testing
phases (since otherwise a lot of rework and retesting has to be expected).
Nonetheless, many projects start late and find themselves confronted
with a backlog of violations that has to be carefully managed
to minimize project delays.
In these cases, prioritization is an important tool to avoid
submerging developers in unmanageable workloads,
and our orthogonal classification may be of help in this regard.
Most importantly, we believe the classification will help projects
with a MISRA compliance objective not to procrastinate the MISRA
effort entirely: it might be acceptable to postpone compliance
with the guidelines dealing with unused/unreachable/dead
code, but other guideline subsets need to be considered earlier.

A good prioritization strategy for such projects is to first make
sure the documentation required to comply with the \DOCU{} guidelines
is available.  Then enforce the \LTLM{} guidelines: among
other things, this will fix the used C dialect and will allow
the compiler qualification efforts to proceed in parallel with
further code development once the compilation options
have been decided.  The \TYPM{} guidelines should be checked next, as
delaying compliance with them typically augments the costs due
to code reworking and refactoring.  The \DEVM{} and \HTDR{} guidelines
should not be
delayed too much if the amount of code still to be written is
substantial.  These are just rules of thumb that the project
manager can adapt to the specifics of the project and the team.

\section{Conclusion}
\label{sec:conclusion}

In this paper, we provide a new, rationale-based classification
of the MISRA~C guidelines that is orthogonal to the classifications
provided in the MISRA documents.
By reminding programmers of the main rationales for violated
guidelines, we believe they can more easily identify the right
course of action in each case.
In particular, the decision whether to comply with or deviate
from a guideline is simplified by the recognition of the
guideline rationale, as this influences both the nature of the
correct remediation measures and the information
to be provided in the deviation record.

A further advantage of the new classification is that it opens
the door to projects without MISRA compliance requirements.
This is the majority of C projects and too many of them are
at the stage where no coding standard is adopted and no
static analysis is performed.
Some of them use bug finders to find some bugs but, as is
well known, bug finders favor false negatives over false positives
and are thus inadequate to implement preventive measures as
in the spirit of the MISRA coding standards.
To projects without MISRA compliance requirements but with
a strong emphasis on code quality and reliability, we propose
adoption of subsets of the MISRA C guidelines that are compatible
with the project objectives and time frame.
This is not in conflict with the possibility of adopting BARR-C:2018
proposed in \cite{BagnaraBH21}: the two approaches are complementary
and both can be applied at the same time or incrementally.

We would like to stress that the notion of ``MISRA compliance''
is formalized in \emph{MISRA Compliance:2020} \cite{MISRA-Compliance-2020}
and is not something open to interpretation.  In particular, there is
no notion of ``partial MISRA compliance'' (the MISRA~C guidelines
constitute a whole that is much superior to any of its parts)
and this paper does not mean to introduce it.
Nonetheless, the availability of good static analyzers\footnote{I.e.,
  roughly: confined false negatives, low rate of false positives,
  small configuration overhead, powerful reporting and filtering
  facilities, easy integration within CI/CD systems.} supporting
the MISRA~C guidelines makes this opportunity valuable for many
projects, especially those that cannot exclude the emergence
of MISRA compliance requirements.

\subsection*{Acknowledgments}

The MISRA~C Guideline headlines are
reproduced with permission of \emph{The MISRA Consortium Limited}.
We are grateful to the following BUGSENG collaborators:
Simone Ballarin, for help with the experimental part of this work;
Lavinia Battaglia, for her precious assistance;
Anna Camerini, for the composition of Figure~\ref{fig:misra-c-history}.
We are also grateful to Frank B{\"u}chner (Hitex GmbH) for a careful
reading of draft versions of this paper, which resulted in several
improvements.

\providecommand{\noopsort}[1]{}

\end{document}